# In-situ crack and keyhole pore detection in laser directed energy deposition through acoustic signal and deep learning

Lequn Chen [a, b], Xiling Yao [c, *], Chaolin Tan [a], Weiyang He [d], Jinlong Su [a], Fei Weng [c], Youxiang Chew [a], Nicholas Poh Huat Ng [b], Seung Ki Moon [b, **]

[a] Advanced Remanufacturing and Technology Centre (ARTC), A*STAR, 3 Cleantech Loop, 637143, Singapore
[b] School of Mechanical and Aerospace Engineering, Nanyang Technological University, 639798, Singapore
[c] Singapore Institute of Manufacturing Technology (SIMTech), A*STAR, 5 Cleantech Loop, 636732, Singapore
[d] School of Electrical and Electronic Engineering, Nanyang Technological University, 639798, Singapore

**Abstract:** Cracks and keyhole pores are detrimental defects in alloys produced by laser directed energy deposition (LDED). Laser-material interaction sound may hold information about underlying complex physical events such as crack propagation and pores formation. However, due to the noisy environment and intricate signal content, acoustic-based monitoring in LDED has received little attention. This paper proposes a novel acoustic-based in-situ defect detection strategy in LDED. The key contribution of this study is to develop an in-situ acoustic signal denoising, feature extraction, and sound classification pipeline that incorporates convolutional neural networks (CNN) for online defect prediction. Microscope images are used to identify locations of the cracks and keyhole pores within a part. The defect locations are spatiotemporally registered with acoustic signal. Various acoustic features corresponding to defect-free regions, cracks, and keyhole pores are extracted and analysed in time-domain, frequency-domain, and time-frequency representations. The CNN model is trained to predict defect occurrences using the Mel-Frequency Cepstral Coefficients (MFCCs) of the laser-material interaction sound. The CNN model is compared to various classic machine learning models trained on the denoised acoustic dataset and raw acoustic dataset. The validation results shows that the CNN model trained on the denoised dataset outperforms others with the highest overall accuracy (89%), keyhole pore prediction accuracy (93%), and AUC-ROC score (98%). Furthermore, the trained CNN model can be deployed into an in-house developed software platform for online quality monitoring. The proposed strategy is the first study to use acoustic signals with deep learning for in-situ defect detection in LDED process.

---

* Corresponding authors:

E-mail addresses: yaox@outlook.com (X. Yao) skmoon@ntu.edu.sg (S.K. Moon)

# 1. Introduction

Laser directed energy deposition (LDED) additive manufacturing (AM) process uses a focused laser beam to melt metallic powders or wires while depositing them on a layer-by-layer basis to form the desired geometry. LDED has gained significant interest in the aerospace, defence, marine and offshore industries over the last decade owing to its unique advantages in fabrication flexibility, waste reduction, surface modification and repair [1]–[4]. In particular, LDED is suitable for producing large metallic parts with higher productivity and lower cost compared to other metal AM techniques, such as laser powder bed fusion (LPBF) and material extrusion [5]. Despite its achievements, LDED still faces substantial challenges in terms of quality consistency and process repeatability. In-situ process monitoring with online anomaly detection is critical to ensure successful AM production [6]; however, it is challenging due to the complicated melt pool dynamics that occur during the rapid melting and solidification process. Many defects (cracking, porosity, layer delamination, etc.) and mechanical properties (hardness, tensile strength, ductility, etc.) can only be observed and evaluated by destructive testing. Most existing non-destructive testing (NDT) methods are still infeasible for online monitoring applications due to the extremely high-temperature environment of the process.

Vision-based in-situ monitoring is one of the most popular monitoring strategies for laser-based AM in recent years. A coaxial vision camera or an infrared (IR) thermal camera can be used to monitor melt pool morphologies and temperature features, which can reflect the melting, cooling, and heat transfer states [7]–[9]. For example, Gonzalez-Val et al. [10] monitored the melt pool during the DED process using a high-speed Medium Wavelength Infrared (MWIR) camera. A CNN model was developed to extract quality indicators from raw images, which were then used to quantify dilution and predict defective spots. Similarly, Grasso et al. [11] monitored the energy-material interactions in selective laser melting (SLM) of zinc powder through infrared imaging. Features were extracted from the IR images, and statistical analyses were conducted to detect unstable melting conditions. Smoqi et al. [12] employed a coaxial pyrometer to obtain thermal images of the melt pool, which were used to extract melt pool signatures such as peak temperature and contour area. The features were fed back into a closed-loop controller, which can improve microstructure homogeneity by reducing localized heat accumulation. A similar approach of vision-based melt pool process control and an adaptive quality enhancement method have also been shown in [11]–[13]. Apart from melt pool monitoring, vision sensors can be used for online surface defect detection [16]–[18], surface roughness, or track geometry prediction of additive manufactured components [19]–[23]. For instance, Li et al. [16] developed a vision-based real-time surface defect (i.e., surface pore, slag inclusions, groove, etc.) detection method through the YOLO algorithm for wire and arc additive manufacturing (WAAM). A surface defect identification approach was recently developed for LDED based on laser line scanning and in-situ point cloud processing [24]. Follow-up research has also demonstrated the capabilities of vision sensors for in-process defect correction [25] for adaptive quality enhancement.

Although vision-based monitoring solutions have attained a certain level of industrial readiness, their implementation is often time-consuming and expensive. Calibration is required for laser displacement sensors or depth cameras to ensure accurate measurement of part surface geometry [26]. The sensing capability of various visual sensors differs significantly as well. For IR thermal cameras, emissivity calibrations are needed to ensure accurate temperature readings. This is especially difficult because metal emissivity varies with temperature, wavelength, material phase, and many other factors [27]. Actual temperature profiles around the melt pool cannot be measured precisely [28]. In addition, the difficulties of sensor integration also limit the use of vision sensors. Coaxial vision sensor installation requires a customized laser head design, while off-axis melt pool monitoring requires image transformation that is less reliable and accurate. For industry end-users, the trade-off between sensing accuracy, sensor prices, and sensor integration complexity is indeed a primary concern.

Acoustic-based monitoring approaches, on the contrary, offer unique advantages such as flexible sensor configurations, fast dynamic response, and cheaper hardware costs. In the LPBF and LDED, acoustic signals produced by laser-material interactions may include information about complicated physical phenomena such as melting, solidification, crack propagation, and pore growth [29]. In addition, the monitoring setup does not require any modification of AM equipment. Such merits make acoustic monitoring particularly attractive to the AM community. Although there is limited study on acoustic monitoring in laser-based AM, it has been extensively used to inspect welding quality, such as penetration depth [30], porosity and cracks [31]. However, since additive manufacturing is a layer-by-layer process with complex geometries, acoustic signal related to defect formation is much more complex than in welding.

Recent research has revealed acoustic-based monitoring approaches in the LPBF process [32]–[35], which has achieved promising outcomes in predicting pore concentrations [36], classifying different materials and defect types (lack-of-fusion, keyhole, balling, etc.) [37], using semi-supervised learning to identify process errors [38], and applying transfer learning to inspects the quality across different materials [39]. The applications were achieved by a low-cost microphone or a fibre Bragg grating sensor, which collected raw acoustic signals and directly used them for training the ML models. In the LPBF process, noise can have a significant impact on the acoustic signal due to the presence of protective gas flow, recoating, and powder delivery systems. Since the laser-powder interactions taking place on a small scale, the noisy chamber environment can affect the acoustic signal for in-situ process monitoring. However, in LPBF, the recoater and powder delivery system are not in motion during the laser scanning, so the noise primarily comes from the protective gas flow. In contrast, LDED has a more complex noise composition, as protective gas flow and the powder stream hitting the substrate are substantial sources of noise, making it difficult to analyse the laser-material interaction sound. Only a few research were reported to tackle the challenge of acoustic monitoring in LDED. For example, Hossain et al. [40] developed a transducer-based sensing device that was mounted to a part's substrate and collected acoustic emission (AE) signals during the DED process. The statistical method was used to verify that AE signals are correlated to the DED part quality. However, the proposed sensor setup lacks flexibility, and further investigation is needed. Recent work presented by Prieto et al. [41] also showed the possibility of using microphone acoustic signal for crack detection in DED, while the investigation was still in the proof-of-concept stage. Similar research on AE monitoring in LDED has been reported by Gaja and Liou [42] and Hauser et al. [43]. While prior studies focused on acoustic signal analysis, feature extractions and monitoring, our study adds a novel aspect to the field by combining acoustic denoising and features with deep learning for defect classification in the LDED process.

To this aim, this paper proposes a novel in-situ defect detection strategy in LDED using a microphone with deep learning. The key novelty of this work is to develop an in-situ acoustic signal denoising, feature extraction, and laser-material interaction sound classification pipeline that incorporates cutting-edge convolutional neural networks (CNN). To enable in-situ monitoring, a Robot Operating System (ROS)-based software platform is developed, which executes the acoustic signal processing and deep learning pipeline and predicts the defect occurrences on-the-fly. An acoustic denoising technique is used to clean the raw acoustic data, which includes noises from machine moving, protective gas flow, and powder flow. Following that, key acoustic signatures are extracted in time-domain, frequency-domain and time-frequency representations. Using the Mel-Frequency Cepstral Coefficients (MFCCs) features of the laser-material interaction sound, the CNN model is trained to differentiate sound from defect-free regimes, crack and keyhole pore regimes. The CNN model is compared to a number of classic machine learning (ML) models trained on denoised and raw acoustic datasets. The validation results show that the CNN model trained on the denoised dataset outperforms others with the highest overall accuracy (89%), keyhole pore prediction accuracy (93%) and ROC-AUC score (98%). The proposed strategy is the first study to use acoustic signals with deep learning for in-situ defect detection in LDED process, which can identify location-specific defects including cracks and keyhole pores.

The rest of the paper is structured as follows. Section 2 provides an overview of the proposed framework for in-situ defect detection using deep learning. Section 3 illustrates the experimental procedures, dataset preparations, software architectures, as well as the proposed acoustic signal denoising technique, extraction of key acoustic features, and training details for the CNN and ML models for defect classification. The results of the model's performance evaluation and validation are discussed in Section 4. Lastly, Section 5 concludes by summarizing the key findings of the research and proposing further work on in-situ acoustic monitoring for the LDED process.

## 2. Deep learning-assisted acoustic-based in-situ defect detection framework

Figure 1 illustrates an overview of the proposed acoustic-based in-situ defect detection framework, which consists of an in-situ acoustic denoising, feature extraction, and laser-material interaction sound classification pipeline. Firstly, a signal denoising technique is applied to clean the noisy LDED sound. Section 3.3 provides details of acoustic signal denoising and its results. Following that, key acoustic signatures in the time-domain, frequency-domain, and time-frequency representations (Cepstral-domain) are extracted from the denoised acoustic signal. Feature correlations and their connections with LDED defects are quantitively investigated and discussed in Section 3.4. Subsequently, a CNN model and various traditional ML models are trained to classify the LDED sound into three categories, including defect-free, cracks, and keyhole pores. The CNN model fed on MFCCs features yielded the best performance (overall accuracy of ~89%) among all models, which was incorporated into the software for online defect detection. Section 3 describes the details about system setups, experimental procedures, dataset descriptions and each steps in the proposed defect detection framework.

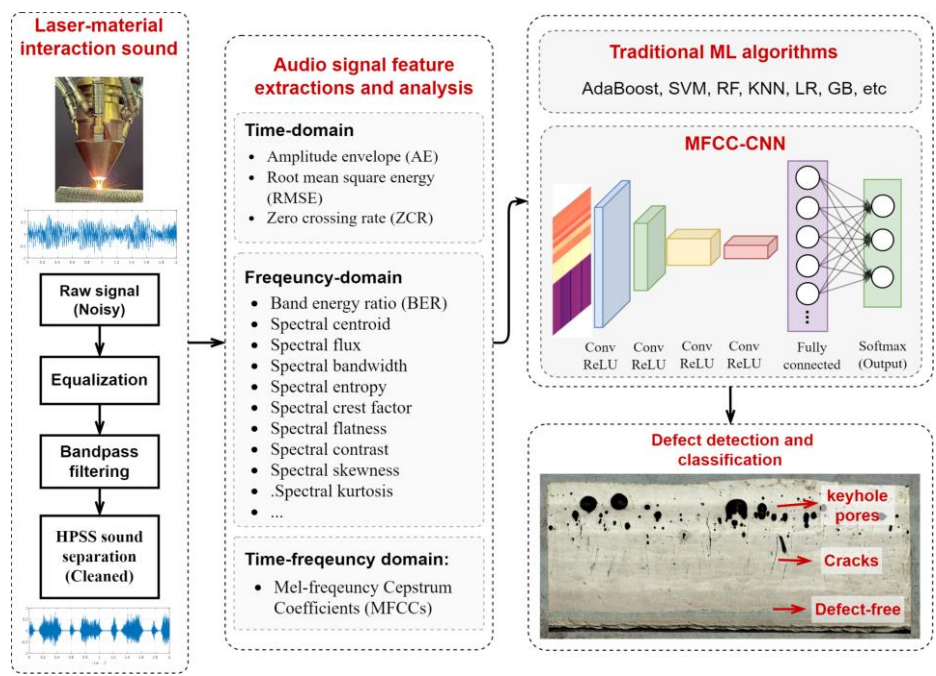

**Figure 1**. Overview of the proposed in-situ defect detection framework through acoustic signal processing and deep learning.

## 3. Methodology
### 3.1 Experimental procedures and raw acoustic datasets

Figure 2 depicts the robotic LDED in-situ acoustic monitoring system used in this study. The system is equipped with a six-axis industrial robot (KUKA KR90) and a two-axis positioner. A laser head and a coaxial powder feeding nozzle are attached to the robot arm's end-effector. The LDED process sounds were recorded using a Prepolarized microphone sensor (Xiris WeldMIC) with a frequency response ranging from 50 to 20,000 Hz. The Prepolarized microphone sensor in this study does not require any external power supplies or preamplifiers. The microphone can be directly connected to the laptop for audio signal processing. The microphone is placed next to the laser head (approximately 10 cm from the molten pool) at an angle around 30 degrees, with 44,100 Hz sampling rate to satisfy Shannon Nyquist theorem [44] (where the analogue signal can be converted to digital and back to analogue without any significant loss of information).

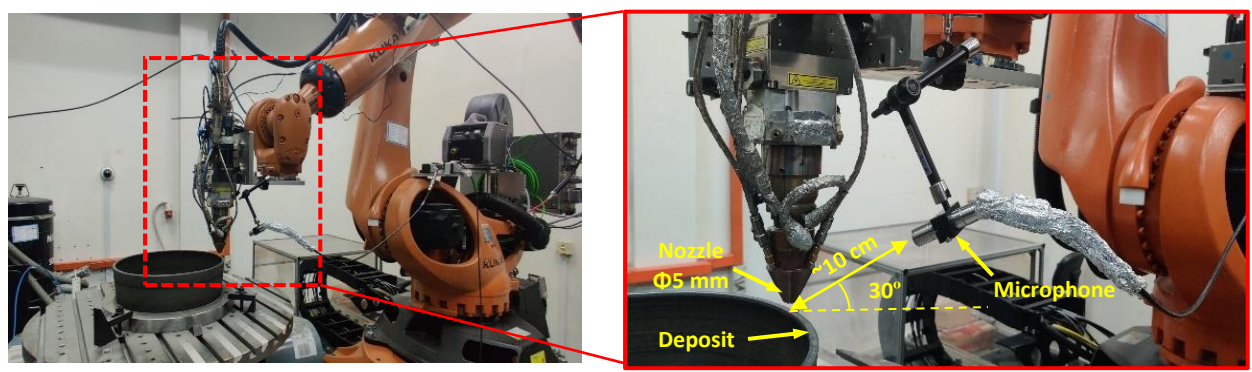

**Figure 2**. In-situ acoustic monitoring setup for the robotic LDED system.

In the powder-blown LDED process, defect occurrences are difficult to forecast because of the dynamic and stochastic nature of the melt pool metallurgical process [45]. Transitions from conduction mode (e.g., non-defective mode) to abnormal states like keyhole mode melting regimes are particularly difficult to determine [46]. In most circumstances, the trial-and-error approach [47] or mechanistic modelling approach [48] is used to obtain optimal process parameters for producing dense and defect-free parts. However, variations in part quality are still seen even when the optimal process parameters are applied. The substrate temperature rises as the process continues, resulting in nonuniform tracks, an extended heat-affected zone, excessive dilution, geometric distortion, and cracking due to residual stress build-up. Furthermore, unstable melt pool dynamics and high energy density may result in material evaporation, which creates keyhole pores. Keyhole pores and cracks are the most severe defects in LDED, which directly degrades mechanical performance, such as strength, microhardness, and fatigue life [46, 47].

In this study, we produced a number of single bead wall structures with varying process parameters using commercial Maraging Steel C300 powder material to create an AM acoustic dataset, as shown in Tables 1 and 2. Unlike most existing sensor-based defect detection research, we do not deliberately use suboptimal process settings to create defects. Instead, we employ pre-optimized process parameters to deposit materials from start to finish, allowing us to see the transition from the defect-free to the defective regime. The process parameters were optimized through trial-and-error experiments with depositing block samples. The original optimized print parameters were used for Experiment #1, as shown in Table 1. When printing the single bead wall samples for acoustic data collection, we kept the energy density (P/v) constant while adjusting the laser power and scanning speed proportionally. The dwell time between each layer was varied in different experiments to postpone the occurrence of defects since it can reduce localized heat accumulation. As a result, defects appeared in different layers for different samples. The sample fabricated with a longer dwell period contained fewer flaws. Cracks and keyhole pores emerged at a higher layer in samples fabricated with longer dwell time. For each sample, optical microscope (OM) images were taken to identify locations of cracks and keyhole pores within the part. Figure 3(a) shows the OM image (x-z outer surface) of a single bead wall sample produced for acoustic data collection. The wire-cutting process removes the outer surface of the single bead wall, allowing us to observe the location-specific quality. As shown in Figure 3(a), the process transitions from the defect-free regime to the crack regime after several layers of deposition due to heat build-up. As the process progresses, significant heat accumulation causes material evaporation and gas entrapment in the molten pool to form keyhole pores in the sample's upper layers. The acoustic signal was segmented to 500 ms pieces and spatiotemporally registered with defect locations for data labelling. During the experiments, the acoustic signal was recorded simultaneously and synchronously with the robot tool-centre-point (TCP) position data through the in-house developed ROS software. If cracks or keyhole pores occurred at a specific location (as observed from the OM image), the 500 ms acoustic signal segment corresponding to that location was marked as "cracks/keyhole pores". This process enabled us to create a labelled dataset for training the defect detection model.

The total acoustic dataset consists of 1300 signal samples segmented at 500 ms length from three categories: defect-free, cracks and keyhole pores (Figure 3(b)). The acoustic signal before and after each step of denoising were also collected, which were used to validate the effectiveness of the proposed denoising approach. Figure 3 (c)-(d) displays LDED sounds after denoising from each category. The keyhole pore has the largest magnitude, and cracks have a distinct amplitude envelope, whereas the defect-free regime has a more stable and smaller amplitude.

Table 1. LDED experiments for acoustic data collection.

| Experiment | Laser power, $P$ (kW) | Speed, $v$ (mm/s) | Dwell time (s) | Powder flow rate, $f$ (g/min) | Energy density, $P/v$ (kW·s/mm) | Line mass, $f/v$ (g/mm) | Types of defects generated |
|---|---|---|---|---|---|---|---|
| #1 | 2.3 | 25 | 0 | 12 | 0.092 | 0.480 | cracks, keyhole pores |
| #2 | 2.53 | 27.5 | 0 | 12 | 0.092 | 0.436 | cracks, keyhole pores |
| #3 | 2.3 | 25 | 5 | 12 | 0.092 | 0.480 | cracks, keyhole pores |
| #4 | 2.3 | 25 | 10 | 12 | 0.092 | 0.480 | cracks, keyhole pores |
| #5 | 2.53 | 27.5 | 5 | 12 | 0.092 | 0.436 | cracks, keyhole pores |
| #6 | 2.53 | 27.5 | 5 | 12 | 0.092 | 0.436 | cracks, keyhole pores |

Table 2. Other LDED process settings during the experiments.

| Parameters | Values |
|---|---|
| Geometry | Single bead wall structure |
| Dimension | 90 mm * 42.5 mm |
| Number of layers for each sample | 50 |
| Laser beam diameter | 2 mm |
| Layer thickness | 0.85 mm |
| Stand-off distance | 12 mm |
| Laser profile | Gaussian |
| Laser wavelength | 1064 nm |
| Material | Maraging Steel C300 |

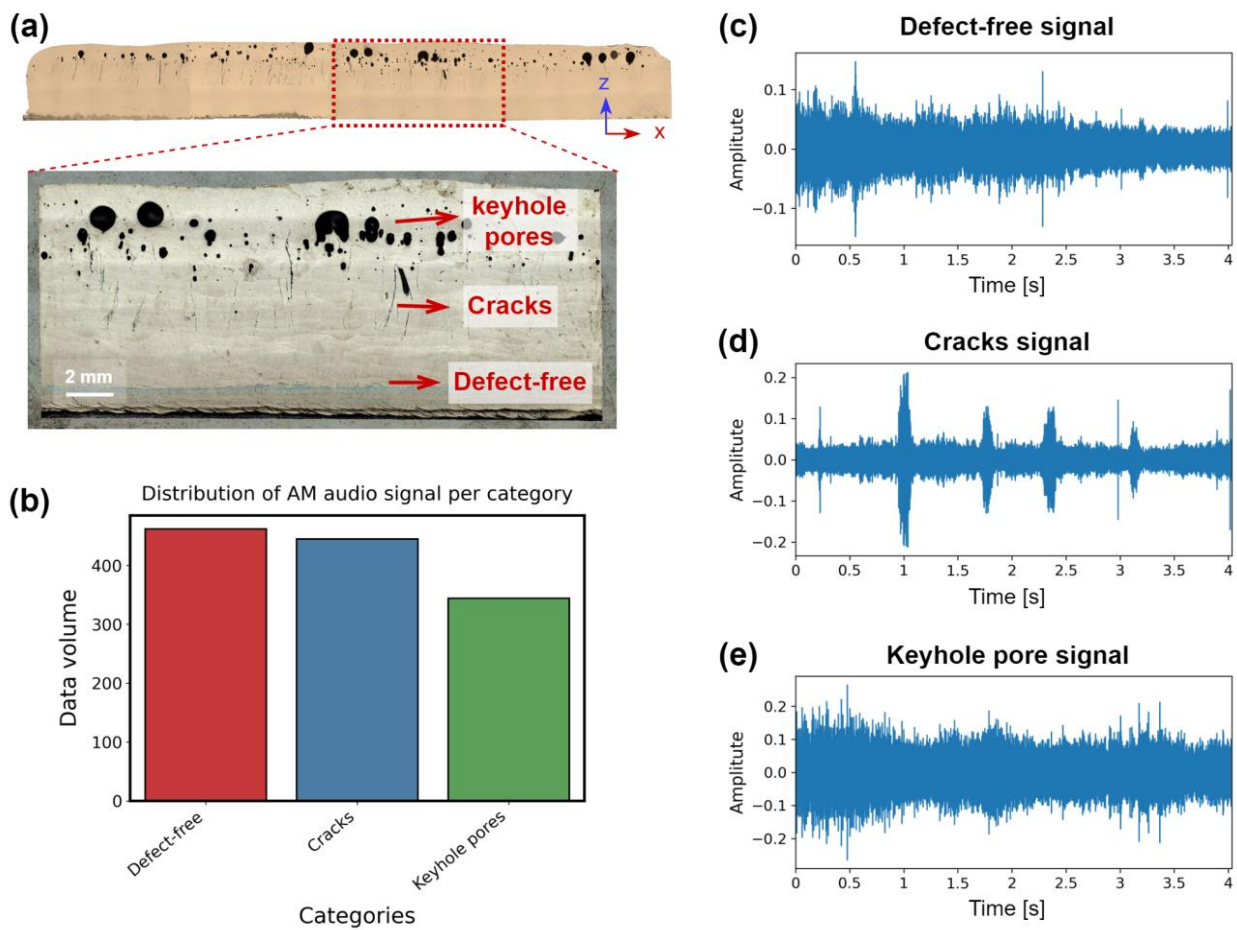

**Figure 3**. LDED audio dataset preparations and descriptions. (a) An OM image of the single bead wall sample produced for acoustic data collection (Image taken from x-z outer surface of Experiment #1 in Table 1. (b) Distribution of the AM audio dataset per category, where acoustic signals from each category were segmented into 500 ms pieces. (c)-(e) Visualization of a 4-second-long acoustic signal piece (one layer) from each category.

## 3.2 Software architecture for acoustic-based defect detection

The in-house designed software is deployed on a personal computer (PC) running Linux (Ubuntu 20.04LS) that works as the central controller for acoustic-based defect identification. The in-situ acoustic monitoring software adopts similar multi-nodal philosophy as the one reported in [51] and [24], where Robot Operating System (ROS) open-source framework [52] is used to establish the communications among the sensor, robot, and PC. Figure 4 shows the proposed software architecture, consisting of ROS nodes for raw acoustic signal capturing, denoising, feature extraction, ML/DL models for defect prediction, and online feature data visualization. The ROS nodes runs simultaneously and data is exchanged over topics channels. "Subscribe" and "Publish" describe the communication mechanism between ROS nodes. "Subscribe" refers to a node receiving data from another node through a topic channel. "Publish" refers to a node sending data to other nodes by publishing it to a topic. Nodes can both subscribe to and publish multiple topics, enabling flexible communication within the robotic system. The use of publish/subscribe messaging model can be found in many event-driven systems and Internet of Things (IoT) platforms, including Message Queuing Telemetry Transport (MQTT) [53], Data Distribution Service (DDS) [54], and Apache Kafka [55], where devices can both publish and subscribe to topics to exchange information. Therefore, platforms other than ROS can employ the same software architecture represented in Figure 4. The details of the software architectures are illustrated below.

- "Microphone sensor capturing node" extracts raw acoustic signal with time stamps and publishes it as a ROS topic. The raw acoustic data is captured at 44100 Hz and stored in a buffer. ROS publishes the time-stamped signal at a frequency of 30 Hz.
- "Acoustic signal denoising node" subscribes to the raw acoustic data and conducts the denoising algorithms (i.e., equalization, bandpass filtering, and Harmonic-Percussive Source Separation (HPSS) [56]). It publishes the time-stamped denoised signal as a ROS topic at a frequency of 30 Hz. Raw signal and the denoised signal can both be subscribed for offline data analysis
- "In-situ feature extraction node" subscribes to the denoised data and extracts key features such as amplitude envelop (AE), spectral descriptors and MFCCs. The acoustic feature extraction was implemented using nussl [57] and librosa [58] library.
- The ML models (e.g., KNN, SVM, gradient boosting, etc.) and MFCC-CNN model were loaded in a ROS node that subscribes to the extracted features and stores them into a buffer. The models can make an inference and publish the predicted defect as a ROS topic each 500 ms.
- All the features can be visualized online via the PlotJuggler plugin [59].

The proposed in-situ defect detection strategy can predict the occurrence of defects while the machine is in operation, as opposed to relying on ex-situ quality inspection. The ML model publishes its predictions to a ROS topic every 500 ms for each segment of the acoustic signal. This feedback signal represents the current quality and can be used for closed-loop process adjustment. The software can issue warnings when the defects are detected, and the process can be stopped immediately to prevent further quality deterioration. Alternatively, the laser power can be reduced when defects are detected, minimizing the localized heat accumulation.

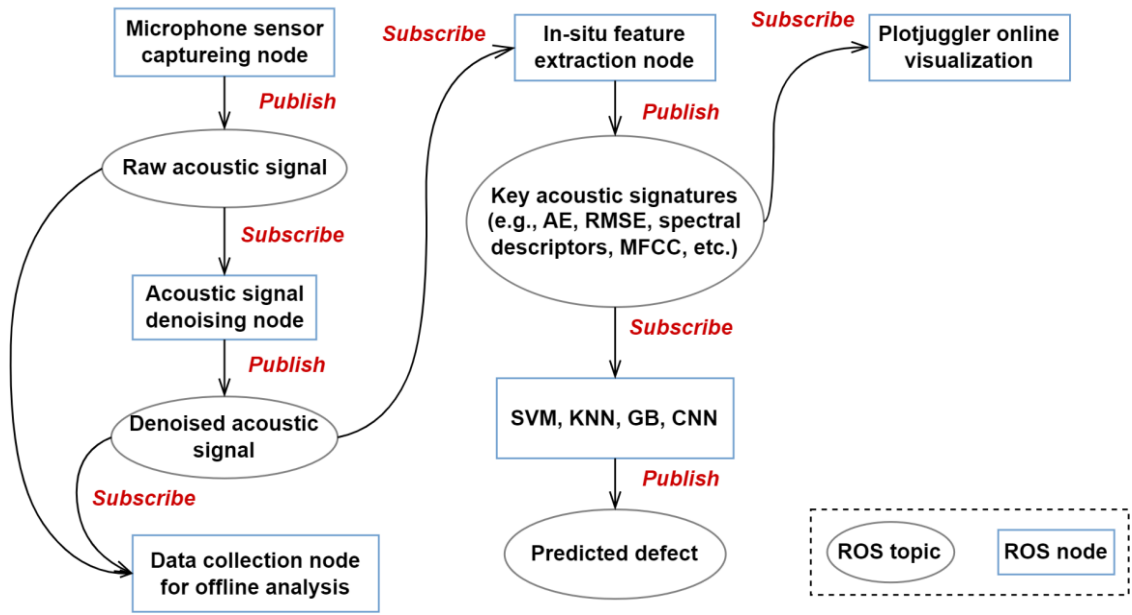

**Figure 4**. In-house developed ROS-based software architecture for in-situ monitoring and defect prediction for the LDED process through a microphone sensor.

## 3.3. Acoustic signal denoising

The LDED process's raw acoustic signal incorporates noise from several sources, including machine motion, powder flow, and protective gas flow. Figure 5 depicts the Fast Fourier Transform (FFT) [60] for various sound sources during the LDED process in different scales. The magnitude is depicted in linear scale in Figure 5 (a). The magnitude is depicted in logarithmic scale "decibels (dB)" in Figure 5 (b). In this scenario, the decibel values are logarithmic of the magnitudes of the normalized audio data samples, allowing us to examine and analyse the structure and content of the audio signal more easily. Figures 5 (a) and (b) compare the FFT frequencies in logarithmic and linear scales, respectively. The FFT plots in Figure 5 (a) reveal signal content in the low frequency band (<1000 Hz), whereas the FFT plots in Figure 5 (b) reveal more information about the audio signal content above 1000 Hz. Furthermore, each sound component was captured separately, with no other source active. The LDED sound in frequency range 0-1000 Hz overlaps the most with the noise content ("Machine + Ar gas + Powder flow"), whereas the amplitude of LDED sound in high-frequency bands above 10 kHz is obviously higher than the individual noise content. As a result, the signal of interest is the signal in high frequency bands, where LDED laser material interaction sound dominates. In addition, machine sound evidently contributes the most to the low-frequency bands noise, whereas other sound sources have relatively little influence on the total audio output. Nevertheless, since all sound sources make a contribution to energy across the entire frequency bands, acoustic signal separation is challenging. A LDED sound source separation technique was proposed in order to isolate the signal of interest from the noisy surroundings in our previous study [61], which are briefly illustrated as follows.

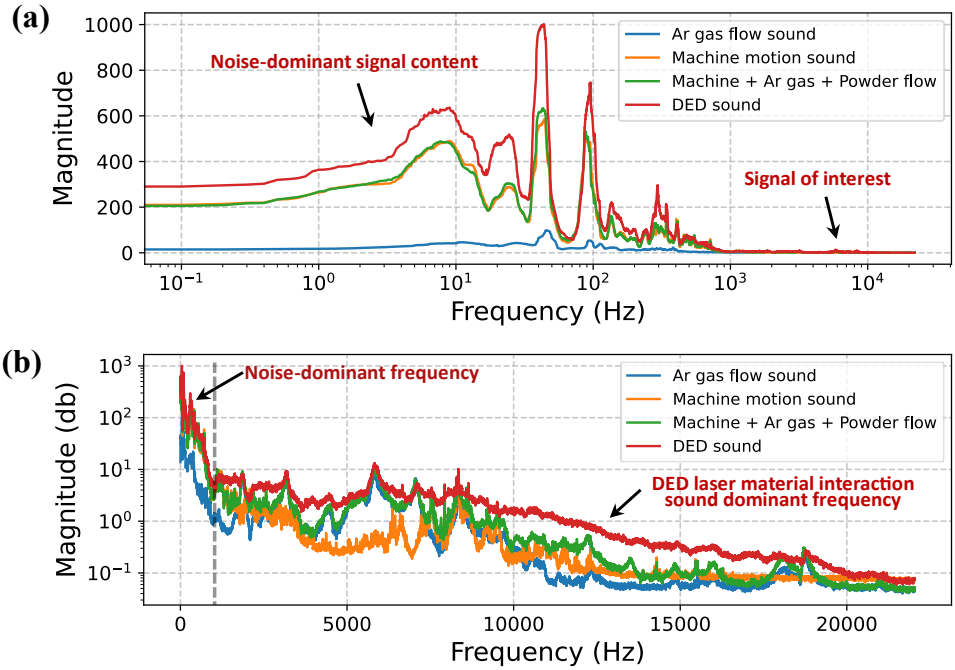

**Figure 5**. Plots of the Fast Fourier Transform (FFT) for various acoustic signal sources throughout the LDED process: (a) frequency in logarithmic scale, magnitude in linear scale; (b) magnitude in logarithmic scale, frequency in linear scale.

The raw signals were first modified using acoustic equalization technique [62], which changes the magnitude of different frequency bands. The transfer function $H_{eq}(z)$ of a parallel acoustic equalization can be written as:

$$H_{eq}(z) = \sum_{m}^{M} G_m H_m \qquad (1)$$

where $H_m$ denotes the transfer function of different frequency band, and $G_m$ denotes the signal gains that regulate the amplitude of each bandwidth. To reduce the noise component and increase the volume of the sound generated by laser-metal processing, the gain in the equalizer is manually tuned. The volume of frequency spanning from 1000 Hz to 20,000 Hz were enhanced, while the volume of the frequency outside this region, where machine noise prevails, were muted. Subsequently, to eliminate high and low-frequency noise, a bandpass filter is utilized. The bandpass filter allows frequencies between 1000 and 21,000 Hz to pass through while attenuates frequencies outside the passband. Bandpass filtering was conducted using Python SciPy library with the filter order set to 3. Finally, the Harmonic-Percussive Source Separation (HPSS) [56] technique is used to extract the percussive component of in the LDED sound.

Figure 6 shows the FFT plots for each stage of the three-step acoustic signal denoising approach. The acoustic equalizer reduced low-frequency noise while increasing amplitude from 1000 to 20k Hz. The laser-material interaction sound was magnified by raising the volume of the signal of interest and decreasing the volume of the noise-dominant region. Following that, the bandpass filter attenuates frequencies outside the passband from 1000 Hz to 21,000 Hz. In the last step, the HPSS algorithm retrieved the percussive sound elements of the audio signals. As a consequence, the majority of the environment noise were eliminated or greatly reduced.

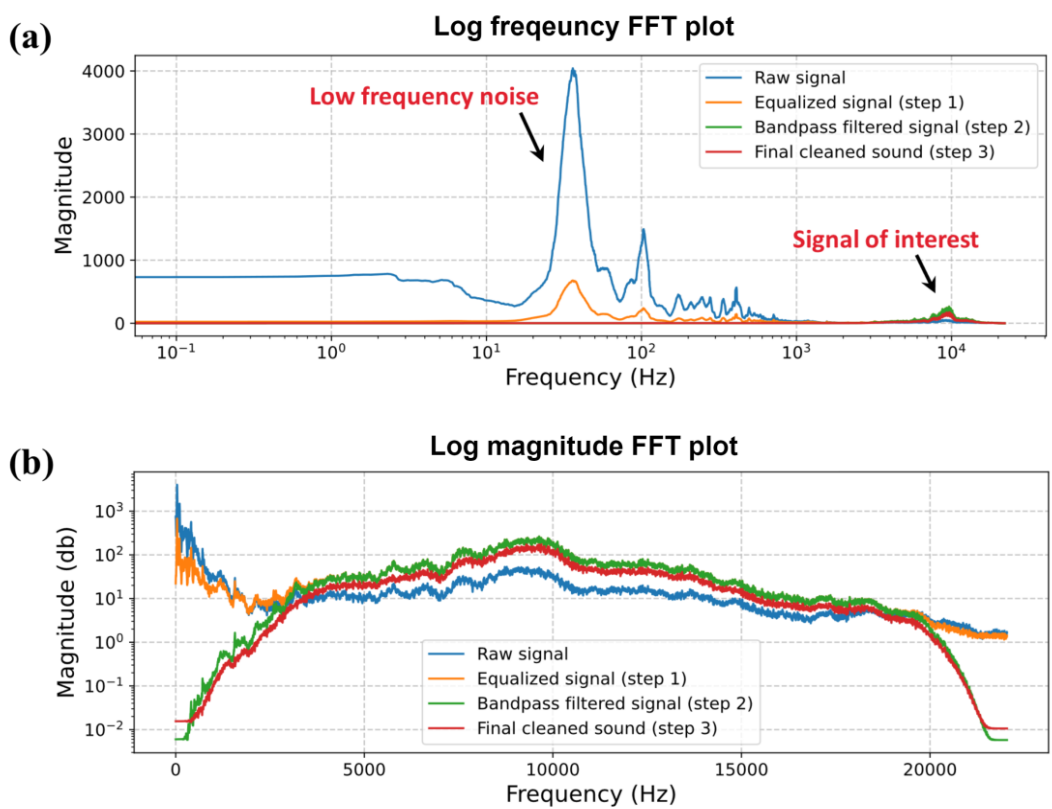

**Figure 6**. Each step of acoustic signal denoising is represented by a Fast Fourier Transform (FFT) plot: (a) frequency in logarithmic scale, magnitude in linear scale; (b) frequency in linear scale, magnitude in logarithmic scale.

The effectiveness of the proposed signal denoising technique is shown in Figure 7. Figure 7(a) compares the raw signal and the final denoised acoustic signal from Experiment #1 (corresponding to the microscope image shown in Figure 3(a)), where the process moves from defect-free regime to crack regime and subsequently to keyhole pore regime. The denoised signal has a more noticeable amplitude envelope. The laser on and off intervals are observable in the plot. An acoustic signal segment from the crack regime was utilized to demonstrate the effectiveness of each denoising step, as shown in Figure 7 (b)-(e). The amplitude envelope of the crack sound is difficult to discern from the raw signal. Following equalization, the signal of interest between 1000 Hz and 20 kHz was amplified, while noise was reduced. The bandpass filtering eliminates any leftover noise, while the final HPSS stage captures the crack sound clearly, as shown in Figure 7(e).

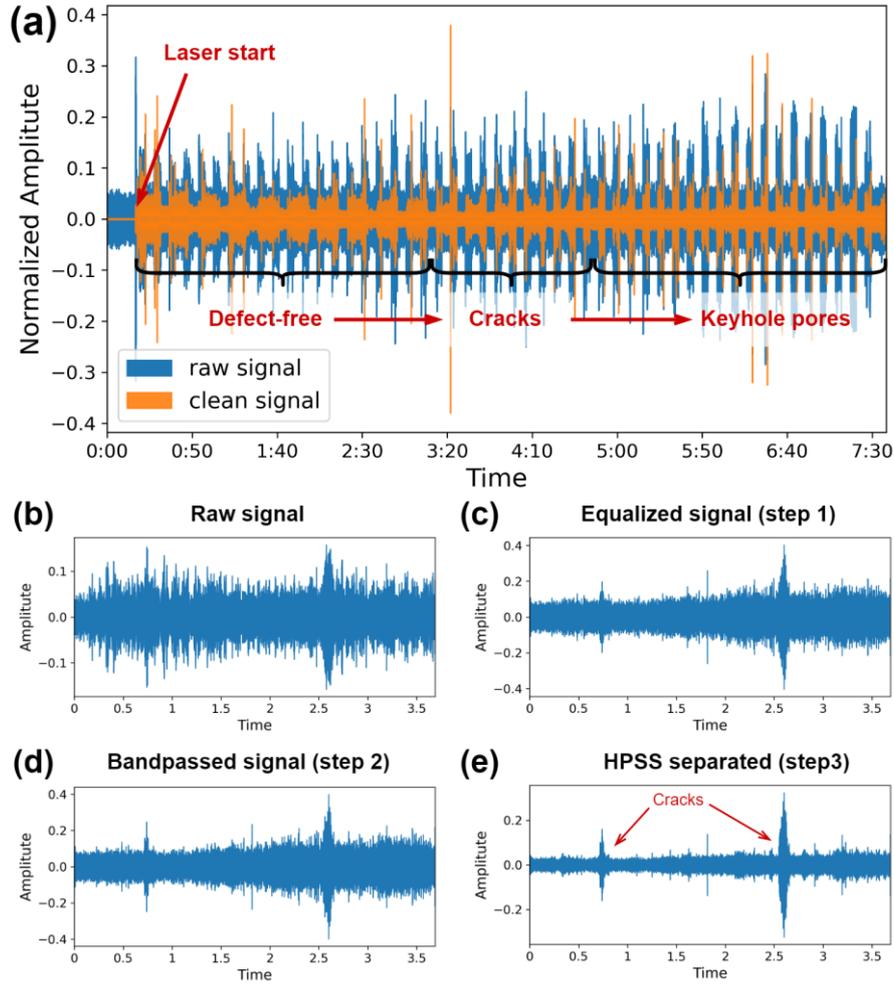

**Figure 7**. Visualization of acoustic signal denoising steps. (a) Comparison of the raw acoustic signal and the final denoised signal from one experiment, corresponding to the microscope image shown in Figure 3(a). The process transitions from the defect-free regime to the crack and then to the keyhole pore regime. (b)-(e) Visualization of the acoustic signal after each denoising step. The selected piece of sound belongs to the crack regime.

## 3.4. Acoustic feature extraction

In this section, key acoustic features in time-domain, frequency-domain and time-frequency representations are analysed. The correlation between acoustic features and the output class (i.e., defect-free, cracks, keyhole pores) is quantitatively investigated. Spearman's formulation [63] was used to compute the correlation ($r_{ij}$) between the acoustic features and the categorical labels:

$$r_{ij} = 1 - \frac{6\sum d_{ij}^2}{n(n^2 - 1)} \qquad (2)$$

where $d_{ij}$ represents the distance between the rankings of the $i^{th}$ and $j^{th}$ feature variables, and $n$ denotes the total number of data points. $r_{ij}$ runs from -1 (denoting the strongest negative correlation) to 1 (indicating the strongest positive correlation), with 0 denoting no correlation. The complete feature correlation matrix is shown in Appendix A. Figure A1. Each feature is discussed as follows.

### 3.4.1 Time-domain features

Table 3 summarizes the time-domain acoustic features, their mathematical definitions and descriptions. Before extracting the time-domain features, the windowing parameters in the librosa audio signal processing library [58] must be specified, with the frame size set to 512 and the hop length set to 256. Windowing [64] is the method of analysing a long audio signal into small pieces of the quasi-stationary signal using a sliding window over time. Three time-domain features were extracted, and the mean and variances of these features were calculated for each audio data segment.

- Amplitude envelope (AE) is constituted of the frame's maximum amplitude value. The AE feature can indicate how acoustic energy fluctuates over time and reflects the magnitudes variations directly. The AE mean and variance exhibit a positive correlation to the output class, as shown in Figure A1, indicating that keyhole pore sound has larger and much more unstable AE values than cracks and defect-free sound.

- Root-mean-square energy (RMSE) is computed by RMS of all samples in a frame. RMSE, like AE but less sensitive to outlier disruptions, can represent the magnitude and fluctuations of sound across time. RMSE mean and variance also show a positive correlation to the defects, as indicated in Figure A1.

- The frequency at which the sign of a signal changes is referred to as the zero crossing rate (ZCR). Its use has been extensively recognized in voice recognition and music information retrieval, where it is an important factor in identifying percussive sounds. [65]. As mentioned in Section 3.3, the LDED process sound was found to be related to the percussive components in the acoustic signal; hence, ZCR potentially correlates to the amount of material melting during the process. Figure A1 shows that the ZCR mean value has a negative correlation with defects, with a higher ZCR value corresponding to fewer defects and more stable melting conditions.

Table 3. List of time-domain acoustic features and mathematical definitions

| Feature name | Mathematical expression | Description | Ref |
|---|---|---|---|
| Amplitude Envelope (AE) | $AE_t = \max(s_{(k)}[t \cdot K, (t+1) \cdot K - 1])$<br>- $AE_t$: AE at $k^{th}$ frame $t$<br>- $s_{(k)}$: amplitude of sample<br>- $K$: number of samples in a frame | A boundary curve that traces the signal's amplitude through time, capturing how energy in the signal changes. | [66] |
| Root-mean-square energy (RMS) | $RMS_t = \sqrt{\dfrac{1}{K} \cdot \sum_{k=t \cdot K}^{(t+1) \cdot K - 1} s_{(k)}^2}$ | RMS of all samples in a frame: indication of loudness | [67] |
| Zero crossing rate (ZCR) | $ZCR_t = \dfrac{1}{2} \cdot \sum_{k=t \cdot K}^{(t+1) \cdot K - 1} |sgn(s_{(k)}) - sgn(s_{(k+1)})|$<br>- $sgn$: sign of function (+1, -1, or 0) | A signal's frequency crosses the time axis: recognition of percussive vs pitched sounds | [65] |

### 3.4.2 Frequency-domain features

Figure 8 depicts the FFT plots of the denoised acoustic signal from three different categories (i.e., defect-free, cracks, and keyhole pores). The magnitude of keyhole pore sound is considerably larger in the low frequency bands (0-5000 Hz), followed by crack sound and defect-free sound. The magnitude of defect-free sound is higher at frequencies ranging from 5000 to 10000 Hz, whereas crack and keyhole pore magnitudes are similar in this frequency range. The distinct patterns in the FFT plots demonstrate the feasibility of extracting various spectral descriptors for the following sound classification task. Table 4 summarizes the frequency-domain acoustic features, their mathematical definitions and descriptions. Each frequency-domain feature is discussed as follows.

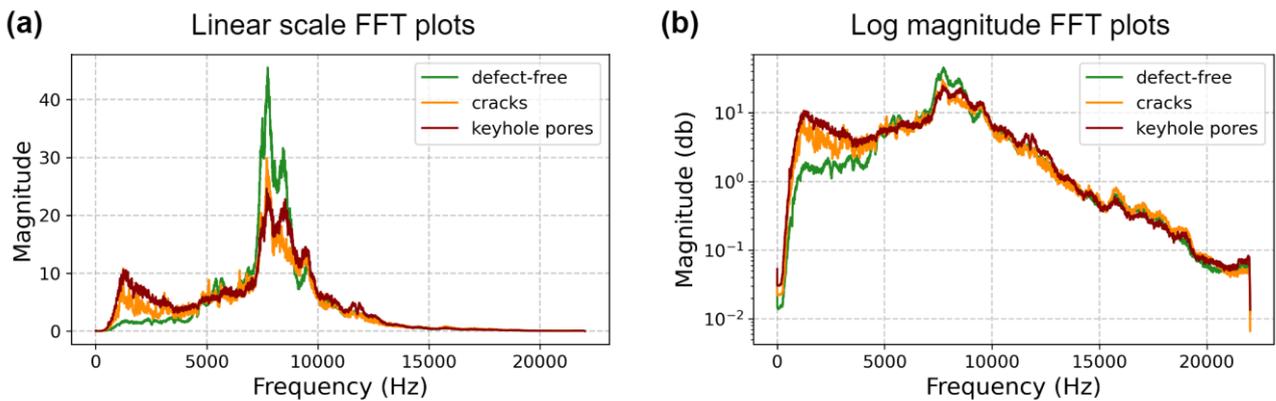

**Figure 8**. Fast Fourier Transform (FFT) plots of the LDED sound from different categories (i.e., defect-free, cracks, and keyhole pores). (a) FFT plot in linear scale, (b) FFT plot in log magnitude scale.

- The spectral centroid (SC) is the centre of gravity (COG) of the magnitude spectrum, which is determined by calculating the weighted mean of all frequencies. Figure A1 shows SC mean value is negatively correlated to defects.

- The spectral bandwidth (SBW) (also known as a spectral spread or dispersion) determines the magnitude spectrum variation from the SC. Figure A1 shows that SBW variation has a clear positive relationship with defects, with keyhole pores resulting in larger variations in SBW. Since SBW can indicate a tone's dominance (e.g., the bandwidth increases as the tones diverge (noise-like) and decreases as the tones converge (rhythms-like)) [68], the finding implies that defect-free sound is more uniform and energy-concentrated, whereas defect sound is more noise-like signal.

- Spectral roll-off (Rolloff) measures the frequency point under which a given percentage (85%) of the total energy exists, and it is often used in music genre classification [69]. As shown in Figure A1, the Rolloff value is also negatively correlated to defects' existence, while the variation of Rolloff is positively correlated to defect.

- Spectral flatness (SF) computes the geometric mean to the arithmetic mean of the power spectrum, which quantifies the frequency distribution's homogeneity. Figure A1 demonstrates that SF negatively correlates to defects.

- Band energy ratio (BER) is defined as the power in low frequency band divided by power in high frequency band. As seen in the previous FFT plots in Figure 8, defects have larger magnitudes in low frequency bands and lower magnitudes in high frequency bands. As a result, the findings in Figure A1 reveal a positive relationship between BER and defects.

- Spectral variance, skewness and kurtosis are the second, third and fourth moments of the spectrum, respectively. These statistical features reflect various magnitude spectrum properties in the frequency domain, which are often employed in music genre categorization and speech recognition [64]. Figure A1 shows that the mean value of variance, skewness and kurtosis all have a negative correlation with the defect existence.

- The spectral crest factor expresses how peaky the spectrum is. It is greater for harmonic sounds and lower for noisy sounds. As shown in Figure A1, it follows the same conclusion as in SBW and SF, where a lower value corresponds to defect sound, which is more of a noise-like signal.

- Spectral entropy is the measure of peakiness and uniformity of energy distribution. As shown in Figure A1, it has a negative correlation with the defects. Spectral flux is the measure of L-2 norm of the spectrum over time, and it is positively correlated to defects.

Table 4. List of frequency-domain acoustic features and corresponding mathematical definitions

| Feature name | Mathematical expression | Description and remarks | Reference |
|---|---|---|---|
| Spectral centroid (SC) | $SC_t = \frac{\sum_{n=1}^{N} m_{t(n)} \cdot n}{\sum_{n=1}^{N} m_{t(n)}}$ | Weighted mean of the frequencies. $n$ represents frequency bands, $m_{t(n)}$ is the spectral value (magnitude) for $n$. $N$ is the range of the frequency bands. | [70] |
| Spectral bandwidth (SBW) | $SBW_t = \frac{\sum_{n=1}^{N} |n - SC_t| \cdot m_{t(n)}}{\sum_{n=1}^{N} m_{t(n)}}$ | Weighted average of frequency band distances from SC (spread of energy). | [70] |
| Spectral roll off (SR) | $SR_t = i \text{ s.t.} \sum_{n=1}^{i} |m_{t(n)}| = \eta \sum_{n=1}^{N} |m_{t(n)}|$ | The central frequency where a particular proportion (85%) of the total energy resides. $\eta$ is the energy threshold (85%). | [71] |
| Spectral flatness (SF) | $SF_t = \frac{(\prod_{n=1}^{N} m_{t(n)})^{\frac{1}{n}}}{\frac{1}{n}\sum_{n=1}^{N} m_{t(n)}}$ | The geometric mean divided by the arithmetic mean of the spectra: determine how much of a sound is noise-like versus tone-like. | [72] |
| Band energy ratio (BER) | $BER_t = \frac{\sum_{n=1}^{F-1} m_{t(n)}^2}{\sum_{n=F}^{N} m_{t(n)}^2}$ | The power in the low frequency band divided by the power in the high frequency band, where $F$ represents split frequency, which was set to 7000 Hz. | [73] |
| Spectral contrast (Contrast) | $Contrast_t = \frac{\sum_{peak} m_{t(n)}^2}{\sum_{valley} m_{t(n)}^2}$ | Taking the mean energy in the top quantile and comparing it to the mean energy in the lowest quantile. High contrast levels are often associated with clear, narrowband signals, and low contrast values are associated with broad-band noise. | [73] |
| Spectral variance ($\mu_2$) | $\mu_2 = \sqrt{\frac{\sum_{n=1}^{N}(n - SC_t)^2 m_{t(n)}}{\sum_{n=1}^{N} m_{t(n)}}}$ | The standard deviation in the vicinity of the spectral centroid. | [74] |
| Spectral skewness ($\mu_3$) | $\mu_3 = \frac{\sum_{n=1}^{N}(n - SC_t)^3 m_{t(n)}}{(\mu_2)^3 \sum_{n=1}^{N} m_{t(n)}}$ | The third-order moment of spectrum, measuring the symmetry around the centroid. | [74] |
| Spectral kurtosis ($\mu_4$) | $\mu_4 = \frac{\sum_{n=1}^{N}(n - SC_t)^4 m_{t(n)}}{(\mu_2)^4 \sum_{n=1}^{N} m_{t(n)}}$ | The fourth-order moment of spectrum. | [74] |
| Spectral crest (Crest) | $Crest = \frac{\max(m_{t(n)}[1, N])}{\frac{1}{N}\sum_{n=1}^{N} m_{t(n)}}$ | The proportion of the spectrum's maximum to its arithmetic mean. | [74] |

| Spectral entropy (H) | $H_t = \dfrac{-\sum_{n=1}^{N} m_t(n) \cdot \log(m_t(n))}{\log(N)}$ | Measures the peakiness of the spectrum. | [75] |
| Spectral flux (Flux) | $Flux_t = \left(\sum_{n=1}^{N} |m_{t(n)} - m_{t-1(n)}|^p\right)^{\frac{1}{p}}$ | Measures variability of spectrum over time, popular in audio segmentation. $p$ is the norm type. $p = 2$ is chosen for L2-norm in this research. | [76] |

### 3.4.3 Time-frequency representations (cepstrum feature)

The preceding investigations solely retrieved acoustic signatures in the time and frequency domains. Time-frequency representations [77] are often more effective approaches for audio signal processing, as the relative energy densities in different frequency bands can be computed. This enables the expression of acoustic signatures in both frequency and time. Common time-frequency representations include spectrogram computed by short-time Fourier transform (STFT) [78], scalogram computed by wavelet transforms (WT) [79], and the cepstrum domain features [80]. In this study, Mel-frequency cepstrum coefficients (MFCCs) [81] from the cepstrum domain were chosen. MFCCs is a common choice for real-time speech recognition applications. The key advantages of MFCCs are their computational efficiency and ability to capture perceptual features. Compared to STFT and WT, MFCC involves fewer computations, making it significantly faster for audio feature extraction tasks. In addition, MFCC can mimic the perceptual sensitivity of the human ear by applying a non-linear transformation of the frequency scale, which approximates the human auditory system's response more closely than the linearly-spaced frequency bands used in the normal spectrum, such as WT and STFT. In our study, MFCCs was chosen as the time-frequency representations because our application was motivated by the fact that a skilled welder may identify defects by listening to the welding sound. AI that has learned from the MFCCs features can therefore achieve human-like performance. Furthermore, software programs for in-situ quality monitoring must be computationally efficient.

MFCCs are determined by taking the inverse Fourier transform of a logarithm of the signal's spectra, which can be represented as follows:

$$\boldsymbol{C}_{(x(t))} = F^{-1}[\log(F[x(t)])] \qquad (3)$$

where function $\boldsymbol{C}_{(x(t))}$ computes the cepstrum of a signal $x(t)$. $F$ represents the Fourier transform function, and $F^{-1}$ is inverse Fourier transform. Figure 9 illustrates a flowchart for practically implementing MFCCs in Python, where the Discrete Cosine Transform is used to reduce the dimensionality for representing the spectrum. Figure 10 shows the results of MFCCs values for a 4-second segment of the denoised acoustic signal from each category (i.e., defect-free, cracks, and keyhole pores). All MFCC values are normalized to a range of -1 to 1. As can be seen, MFCCs is a powerful feature capable of distinguishing the LDED sound from different processing regimes. In the defect-free deposition process, the MFCCs value in low-frequency bands is lower. The brighter colour (value near 1) in cracks and keyhole pores indicates a larger concentration of energy in the low-frequency bands. Due to the fact that cracking is an energy-releasing process, sound waves can readily distinguish such abnormal phenomena by showing unique patterns that reflect the abrupt increase in acoustic energy induced by crack propagation.

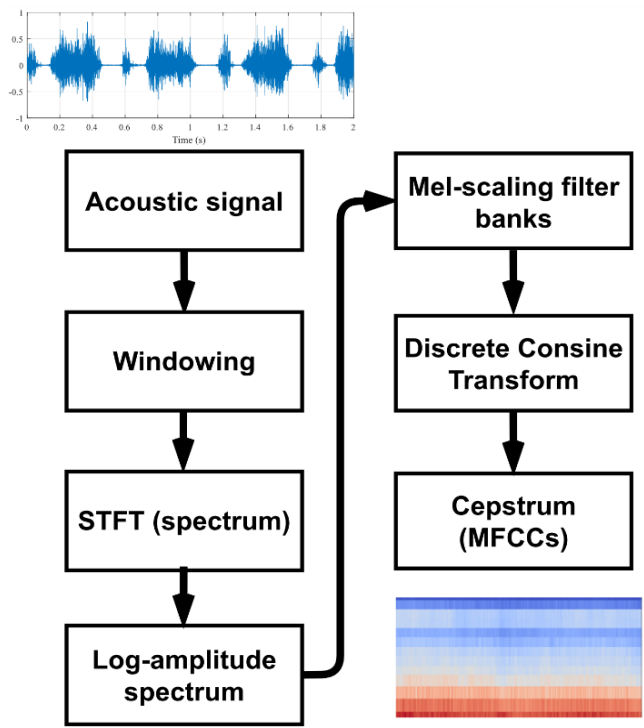

**Figure 9**. Mel-Frequency Cepstral Coefficients (MFCCs) extraction procedure.

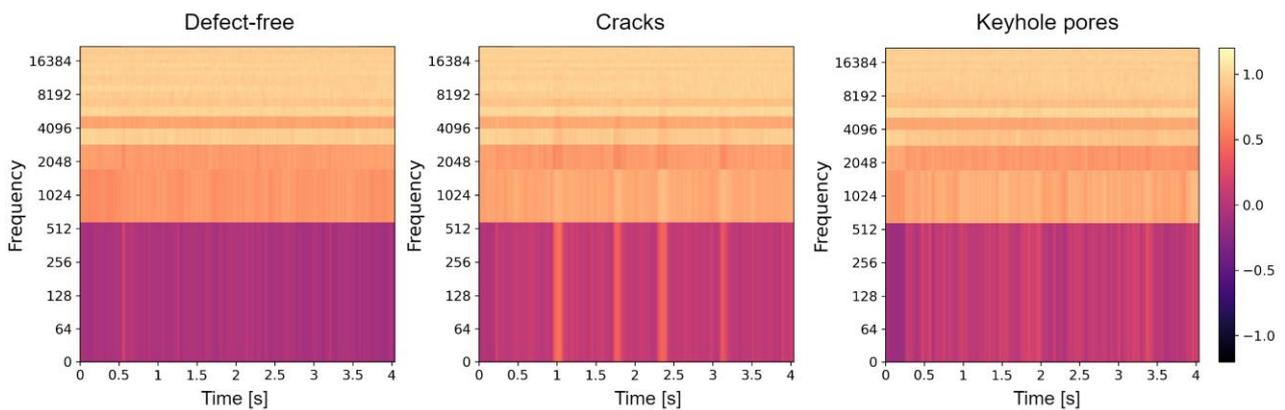

**Figure 10**. Visualization of MFCCs features from each category.

### 3.4.4 Acoustic feature analysis

Finding relationships between acoustic features before putting them into ML models for defect classification tasks can assist select ML model complexity. A feature importance analysis is conducted to determine which features are most important in distinguishing the process regimes (i.e., defect-free, cracks, and keyhole pores). Figure 11(a) depicts the results of a random forest feature importance analysis. The most important features of LDED sound are BER, spectral centroid, entropy, bandwidth, flatness, Rolloff, and variance. However, it is evident that all of the features have a low importance level (with the highest one only slightly larger than 0.12). This is due to the fact that the formation of defects is a highly complex process. None of the individual characteristics could adequately characterize the acoustic signal. Spearman's correlation matrix of the most important acoustic features from the denoised signal is plotted in Figure 11(b). Furthermore, to visualize the high-dimensional acoustic data, we use principal component analysis (PCA) to conduct dimensionality reduction. Figure 11 (c) and (d) show the PCA projection of the raw acoustic signal features and the denoised acoustic signal features, respectively. The results show that denoised acoustic features can form different clusters in low dimensional space, while the raw signal is much more difficult to distinguish.

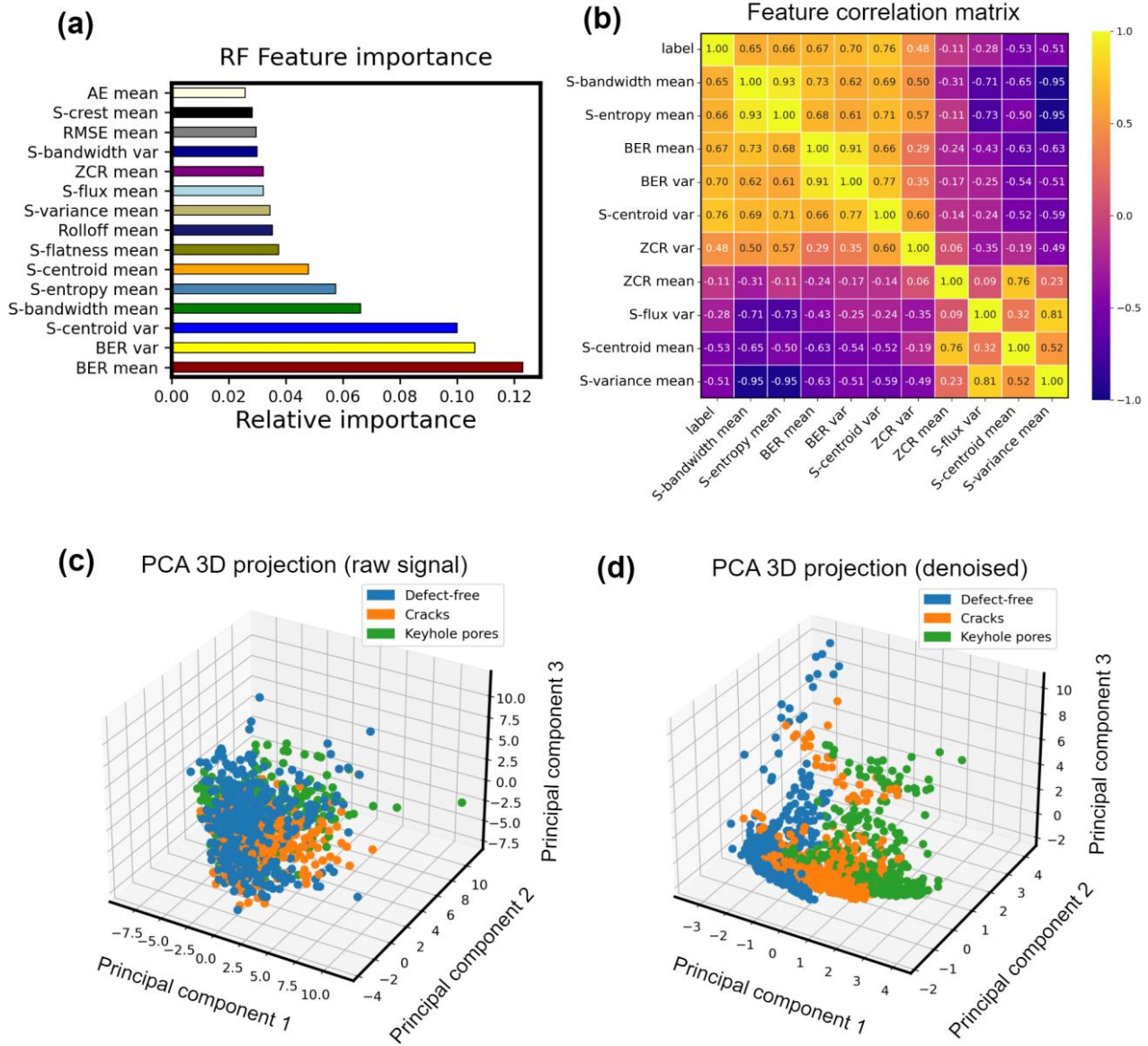

**Figure 11**. Acoustic feature analysis. (a) Random forest feature importance of the denoised acoustic signal. (b) Correlation matrix heatmap of key acoustic features and output class (i.e., defect-free, cracks, and keyhole pores). (c) Low-dimensional feature visualization by PCA projection of raw acoustic signal features. (d) Low-dimensional feature visualization by PCA projection of denoised acoustic signal features.

## 3.5. Defect prediction models

### 3.5.1 MFCC-CNN

A convolutional neural network (CNN) was developed and implemented using the TensorFlow [82] Python Deep learning framework. The CNN model uses MFCCs as input features, as shown in Figure 12. Therefore, it is termed MFCC-CNN. The proposed MFCC-CNN consists of three convolutional layers, a flattened layer, a fully-connected layer and a SoftMax layer. The MFCC-CNN takes the MFCCs values extracted from each segment of the acoustic signal as input, with cepstrum domain features expressed in 20 frequency bands. Prior to being fed into the model, the input features were normalized to have a zero mean and unit variance. For the convolutional layers, the 2D convolution is followed by a ReLU [83] activation function, subsequently, a max-pooling layer. Each max-pooling operation reduces the spatial dimensions of the 2D convolutional layer, and the respective dimensions are shown in Figure 12. The output layer is a SoftMax [84] function which predicts the probability distribution of three output classes (i.e., defect-free, cracks, and keyhole pores). A list of MFCC-CNN model hyperparameters is shown in Table 5. The hyperparameters were optimized by using the k-fold cross-validation grid search method. Adam Optimizer [85] was selected as the optimization solver, which trains the model through minimizing the cross-entropy loss between ground-truths and model predictions [86]. The training was conducted using NVIDIA GeForce RTX 3070 GPU with Keras & TensorFlow Python DL framework. The model performance evaluation will be discussed in Section 4.

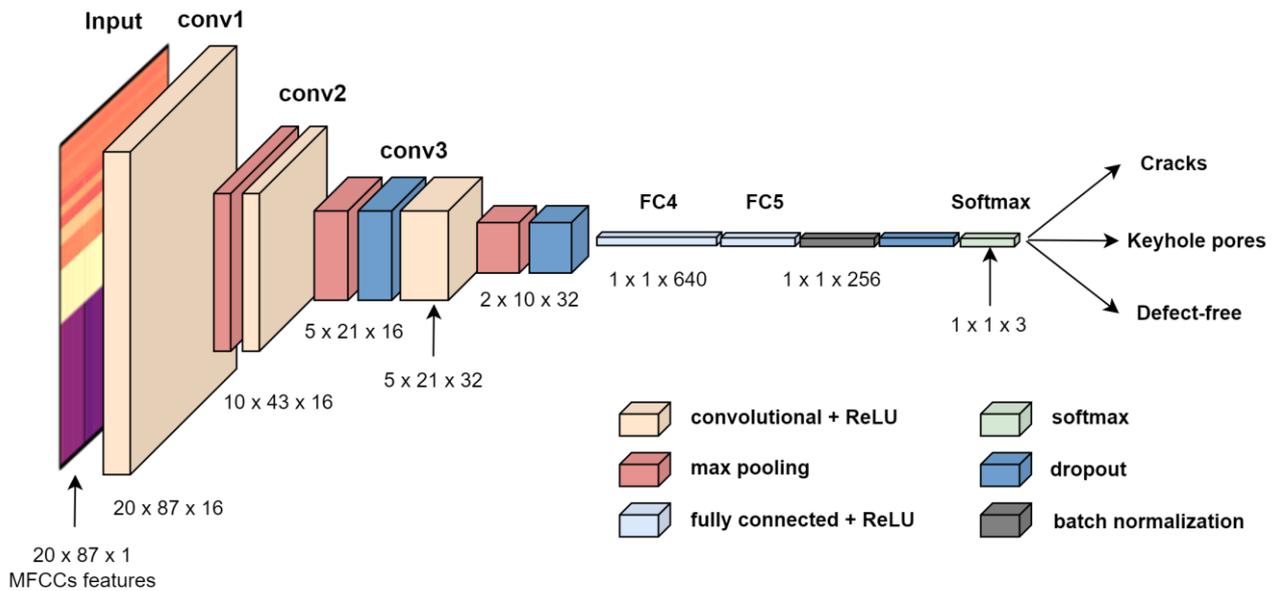

**Figure 12**. The architecture of the Mel-Frequency Cepstral Coefficients-based Convolutional Neural Network (MFCC-CNN).

**Table 5**. MFCC-CNN hyperparameter tuning information

| Training hyperparameter | Optimal values | Range studied |
| --- | --- | --- |
| Solver name | Adam optimizer | ['Adam', 'SGD'] |
| Learning rate | 0.0001 | [1e-2, 1e-3, 1e-4] |
| L2 regularization factor | 0.1 | [0.01, 0.05, 0.1, 0.15] |
| Activation function | ReLU | ['ReLU', 'Sigmoid'] |
| Batch normalization | True | ['True', 'False'] |
| Kernel size in convolutional layer 1-3 | [2, 2, 3] | [[2, 3, 5], [2, 3, 5], [2, 3, 5]] |
| Number of filters (channels) in convolutional layer 1-3 | [16, 16, 32] | [[16, 32], [16, 32, 64], [32, 64, 128]] |
| Stride in convolutional layer 1-3 | [1,1,1] | [[1, 2], [1, 2], [1, 2]] |
| Padding | 'same' | ['same', 'zero'] |
| Number of neurons in the FC5 | 256 | [64, 128, 256] |
| Dropout – conv3 | 0.2 | [0.1 - 0.5] |
| Dropout – FC4 | 0.5 | [0.1 - 0.5] |
| Dropout – FC5 | 0.2 | [0.1 - 0.5] |
| Hop length | 256 | [128, 256, 512] |
| Frame size | 512 | [128, 256, 512, 1024] |

### 3.5.2 Traditional ML models

In this research, we compare the proposed MFCC-CNN model with eight traditional supervised learning algorithms: Naive Bayes (NB), Random Forest (RF), AdaBoost (AB), Decision Tree (DT), Support Vector Machine (SVM), Logistic regression (LR), Gradient Boosting (GB), and K-Nearest Neighbours (KNN). The Scikit-learn Python package [87] was used to implement the ML algorithms for training and testing. The traditional ML algorithms classify LDED sound using the time- and frequency-domain features described in Section 3.4. The input features were selected based on the analysis in Section 3.4.4, including "S-bandwidth mean", "S-entropy mean", "BER mean", "BER var", "S-centroid var", "ZCR var", "ZCR mean", "S-flux var", "S-centroid mean", "S-variance mean".

To optimize hyperparameters for ML models, a grid search approach is utilized, which is an exhaustive search strategy that evaluates all feasible hyperparameter value combinations. Each iteration of the hyperparameter tuning procedure was evaluated using k-fold cross-validation (k=5). The k-fold cross-validation procedure is repeated for each k-fold. The hyperparameter combination with the best cross-validation result is picked at the end of the grid search. The optimal hyperparameter results for the traditional ML models are listed in Table 6.

Table 6. Hyperparameter optimization results of the traditional ML algorithms

| Classifiers | Hyperparameters | Optimal values | Range studied |
|---|---|---|---|
| NB | Variance smoothing | 1e-9 | [1e-10, 1e-9, 1e-8] |
| RF | Minimum split | 3 | [2 - 6] |
| | Number of estimators | 10 | [2 - 10] |
| | Splitting algorithm | Gini impurity | ['Gini', 'entropy'] |
| | Maximum depth | 4 | [2 - 6] |
| AdaBoost | Number of estimators | 10 | [1 - 10] |
| | Algorithm | SAMME | ['SAMME', 'SAMME.R'] |
| KNN | Neighbours | 4 | [3 - 9] |
| | Weight function in prediction | Distance | ['uniform', 'distance'] |
| | Computation of nearest neighbours | Ball Tree | ['auto', 'ball tree', 'kd_tree'] |
| LR | solver | 'lbfgs' | ['lbfgs', 'liblinear', 'newton-cg'] |
| | Penalty | L2 regularization | ['l1', 'l2', 'elasticnet'] |
| SVM | Kernel type | Radial basis function | ['linear', 'poly', 'rbf', 'sigmoid'] |
| | Regularization parameter (C) | 1000 | [1, 10, 100, 1000, 1500, 2000] |
| | Kernel coefficient ($\gamma$) | 0.001 | [1e-2, 1e-3, 1e-4] |
| DT | Minimum samples required to split | 3 | [1 - 10] |
| | Measurement the quality of split | Gini | ['Gini', 'entropy'] |
| | Maximum depth | 6 | [1 - 30] |
| GB | Number of estimators | 10 | [1, 5, 10, 20, 50, 100] |

## 4. Results and discussions

To validate the effectiveness of the proposed denoising technique, the MFCC-CNN and traditional ML models were trained on acoustic signals from different denoising stages. Each denoising step's acoustic signal (raw signal, equalized signal, bandpassed signal, and final denoised signal) was separated into a training set and a testing set for assessing model performance. The ratio of train to test is 8:2. The size of the training dataset is 1080 samples. Since the quantity of data points in each category varies (as shown in Figure 3(b)), the "Stratified Shuffle Split" method in Scikit-Learn was used to create the train and test sets while maintaining the percentage of samples in each class.

The testing accuracy curves for MFCC-CNN trained on the raw acoustic dataset and denoised acoustic dataset are shown in Figures 13 (a) and (b), respectively. The MFCC-CNN trained on the denoised dataset achieves faster convergence and higher testing accuracy, confirming the effectiveness of the proposed acoustic denoising approach. The detailed comparisons are presented and discussed below.

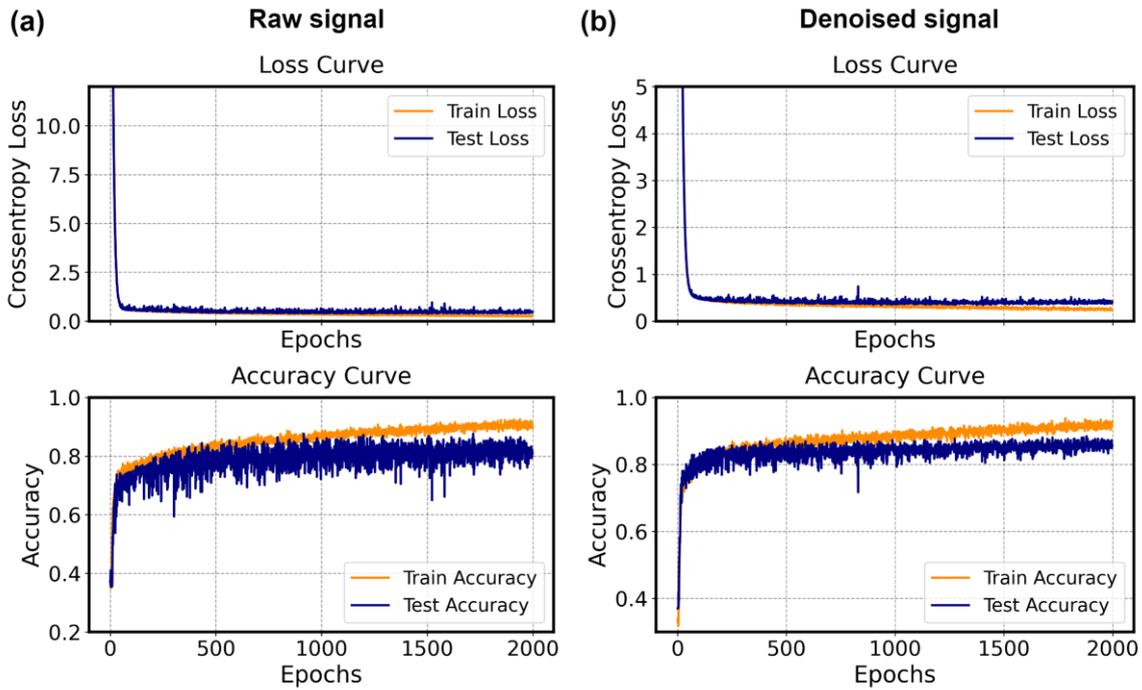

**Figure 13**. Loss and accuracy curves showing CNN models trained on (a) raw acoustic signal and (b) denoised signal. The MFCC-CNN trained on the denoised dataset shows faster convergence and higher testing accuracy.

The proposed MFCC-CNN model and the eight traditional ML algorithms are evaluated in terms of overall classification accuracy, Aera Under Curve of Receiver Operating Characteristics (AUC-ROC) scores, false positive rate (i.e., percentage of actual defects misclassified as 'defect-free' category), and the keyhole pore prediction accuracy. To demonstrate its viability and repeatability, all of the ML model evaluations reported in this paper were averaged over five runs, with standard deviations marked as error bars.

The overall accuracy is the number of correct predictions divided by total predictions, as represented in the following expression:

$$Accuracy = \frac{\#Correctly\ predicted\ samples}{\#Total\ predictions} \quad (4)$$

Figure 14(a) illustrates the classification accuracy of the eight traditional ML models and the MFCC-CNN model trained on the acoustic signal from different denoising phases. In general, the accuracy improves after each denoising step. The AUC-ROC score in Figure 14(b) also demonstrates that, with a few exceptions, such as LR, SVM, and GB, the performance rises with each denoising step. Among all classifiers, the MFCC-CNN model trained on the denoised acoustic dataset had the highest overall prediction accuracy (89%) and the highest AUC-ROC score (98%), confirming the effectiveness of the proposed acoustic denoising technique.

Figures 15 and 16 show the ROC curves and confusion matrix for the different classifiers, respectively. The ROC curves of the MFCC-CNN trained on the denoised dataset outperformed the other models, exhibiting higher AUC values for all predicted classes. Furthermore, the confusion matrix of MFCC-CNN trained on denoised data demonstrates very high classification accuracy on the 'defect-free' class (91.4 %) and the 'keyhole pore' class (92.8 %). Although it does not predict cracks well, the misclassified crack sound is often wrongly labelled as 'keyhole pore', which has little effect on the practical application since both categories are defect sounds.

In this research, the false positive rate (i.e., the proportion of actual defects misclassified as 'defect-free' category) is an essential performance metric. Misclassifying actual defects (such as cracks and keyhole pores) as "defect-free" is detrimental since the system would assume the process could continue without interruption or correction. As a result, the false positive rate should be kept to a minimum. Misidentifying keyhole pores for cracks, on the other hand, has fewer negative consequences because both are defects that must be corrected. Furthermore, false negative decisions (i.e., misclassifying a 'defect-free' regime as defective) are also less harmful because they do not influence the part quality and only affect productivity (i.e., more time spent on process intermittence). Based on Figure 14(c) and (d), the MFCC-CNN model trained on the denoised dataset has the lowest false positive rate (9%), and the best keyhole pore prediction accuracy (92.8%) among all the models.

In addition, a 500 ms prediction window with an overall quality prediction accuracy of 89% was found to be appropriate for our specific application in LDED process. It is worth mentioning that the segmentation length of 500 ms was chosen as a balance between the accuracy of the ML model and the spatiotemporal resolution of predictions. Generally, the accuracy of the ML model decreases as the length of acoustic signals shortens. Longer signals provide more information about the acoustic event, improving the ML model's accuracy. Our choice of 500 ms provides sufficient information for the ML model to make accurate prediction, while still offering a relatively high spatiotemporal resolution. This trade-off was also reported in the study by Tempelman et al. [34]. With our in-house-developed software platform, the defect detection model collects 500 ms of audio samples and infers the presence of defects every 500 ms, publishing the results as a ROS topic. If a keyhole pore or crack is detected within this period, the process can be stopped immediately to prevent further deterioration. A shorter segmentation length would result in a drop in accuracy, which is not desirable.

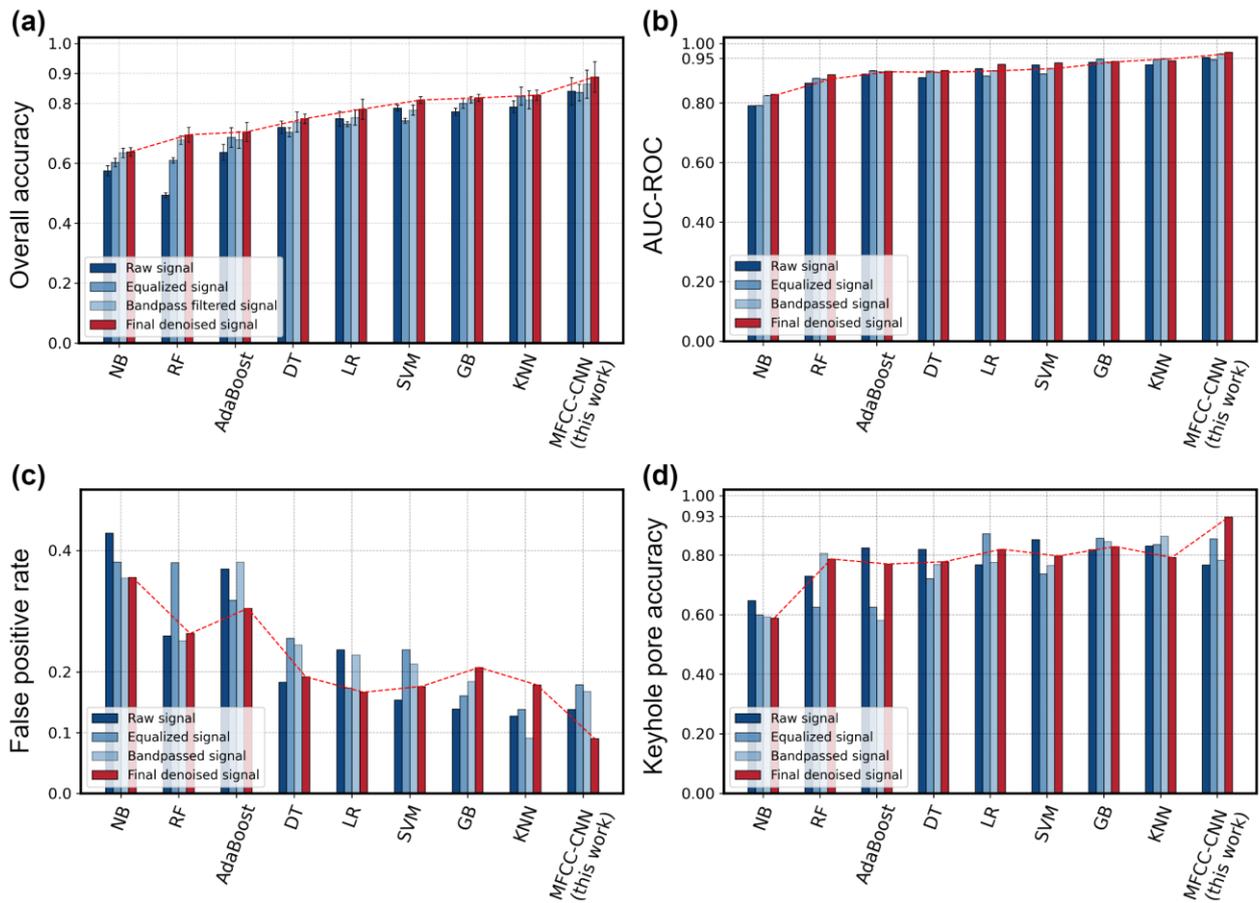

**Figure 14**. Performance evaluation and benchmark for acoustic-based defect prediction in LDED. (a) The overall accuracy of the eight traditional ML models and MFCC-CNN model trained on the acoustic signal from different denoising steps. (b) AUC-ROC results of the eight ML models and MFCC-CNN model trained on the acoustic signal from different denoising steps. (c) False positive rate (i.e., percentage of actual defects misclassified as "defect-free" category). (d) Keyhole pore prediction accuracy. (Note: higher values for overall accuracy, AUC-ROC score, and keyhole prediction accuracy imply better performance; lower values for false positive rate indicate better performance.)

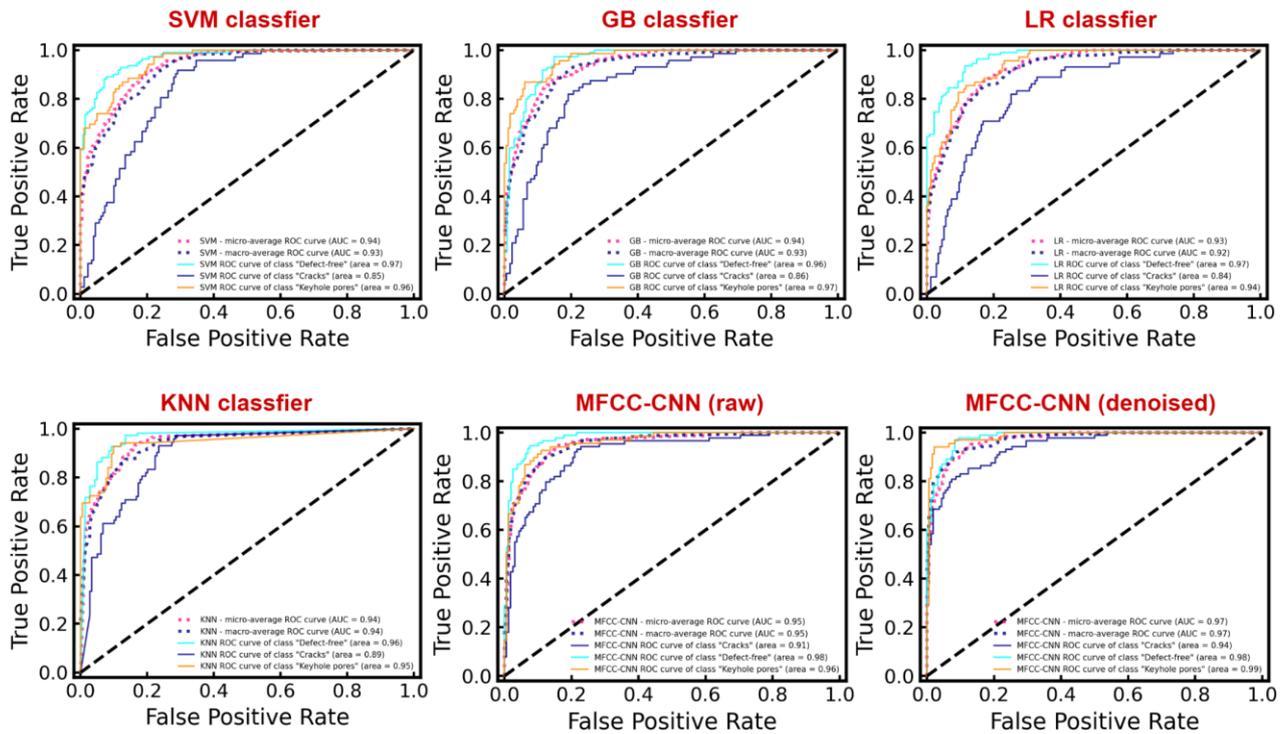

**Figure 15.** Receiver operating characteristic (ROC) curves for prediction of LDED sound by 'Support Vector Machine', 'Gradient Boosting', 'Logistic Regression', 'K Nearest Neighbour', and MFCC-CNN trained on the raw acoustic dataset and denoised dataset. The results shown for SVM, GB, LR, and KNN are trained using denoised acoustic dataset.

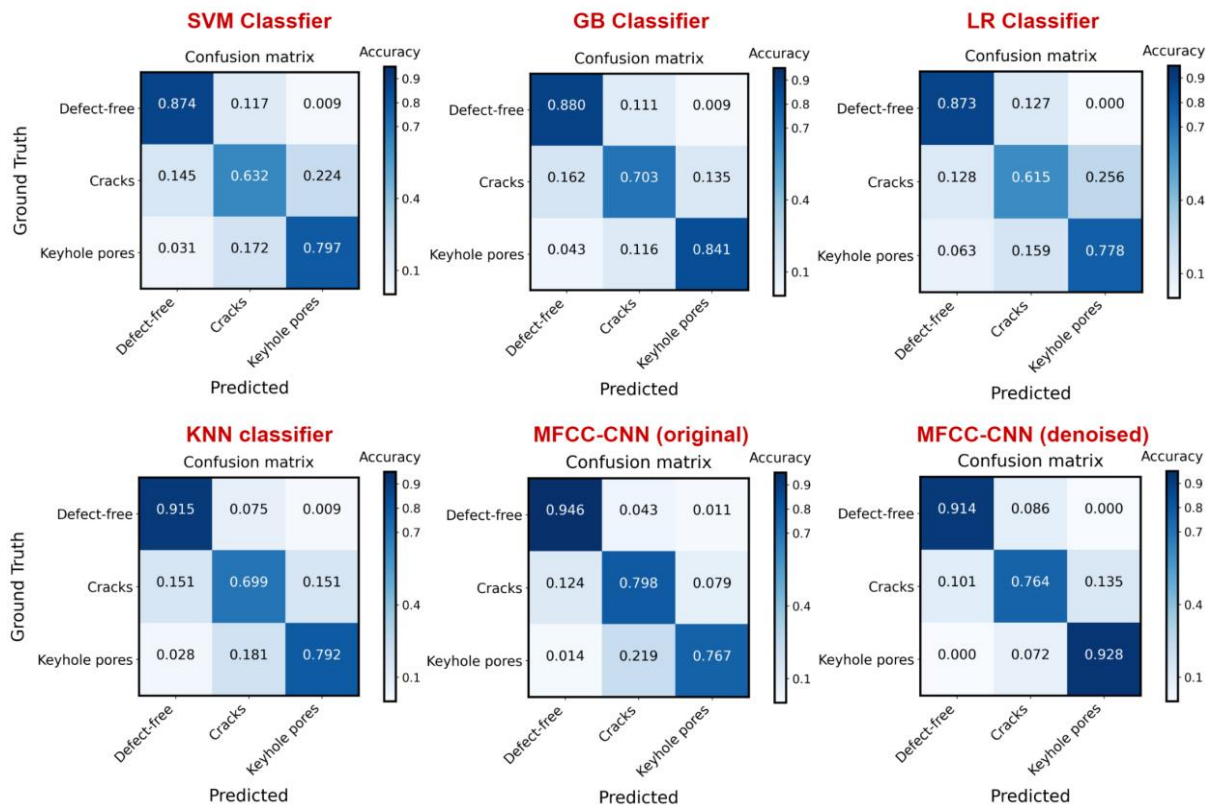

**Figure 16.** Confusion matrix for the classification task for 'Support Vector Machine', 'Gradient Boosting', 'Logistic Regression', 'K Nearest Neighbour', and MFCC-CNN trained on the raw acoustic dataset and denoised dataset. The results shown for SVM, GB, LR, and KNN are trained using denoised acoustic dataset.

## 5. Conclusion and future works

This paper addressed two major challenges in in-situ acoustic-based defect detection for the LDED process: the presence of noise in the LDED laser-material interaction sound and the lack of an automated online defect detection pipeline with in-situ feature extraction and prediction. It was the first study using acoustic signal processing and deep learning for in-situ defect detection in the LDED process. The main contribution and novelty of this work are summarized as follows:

- An automated in-situ acoustic denoising, feature extraction and laser-material interaction sound classification pipeline to predict cracks and keyhole pores in the LDED process.

- A convolutional neural network (CNN) based on Mel-frequency Cepstrum Coefficients (MFCCs) acoustic features to classify LDED sound and predict defects with high accuracy (89%).

- Development of an acoustic signal denoising technique combining acoustic equalization, bandpass filtering, and HPSS algorithm that significantly improves the sound classification accuracy.

- Investigation of key acoustic features corresponding to defect-free, cracks and keyhole pores in the time-domain (e.g., amplitude envelope, RMS energy, etc), frequency-domain (spectral centroid, spectral bandwidth, band energy ratio, etc.), and time-frequency representations (MFCCs).

The proposed MFCC-CNN model surpassed all classic machine learning algorithms that have been tested in this work in terms of classification accuracy, AUC-ROC score, and false positive rate. Furthermore, the model evaluation result demonstrated that the denoised acoustic signal can improve the accuracy and reduce the false positive rate of the sound classification model over the raw acoustic signal. The proposed in-situ defect detection strategy based on acoustic signals and deep learning provides a cost-effective solution for LDED quality assurance by leveraging the flexible microphone setup and lower hardware cost compared to existing sensing methods. However, the timescale of acoustic-based in-situ defect detection in this study is limited (500 ms), which is significantly larger than the work provided in the LPBF process [34] (2.5 ms). On the one hand, the laser scanning speed in LDED is substantially slower than in LPBF. The formation of defects in LDED, on the other hand, is much more challenging to predict because of the noisy environment, making it necessary to have a sufficient length of audio data for the ML model to make an accurate prediction. Future studies will focus on shortening the defect detection period while preserving accuracy. The proposed acoustic-based defect detection framework will also be applied to other alloys and other types of defects such as lack-of-fusion (LoF) and delamination, each of which has a unique acoustic signature and defect formation mechanism. Furthermore, the proposed defect detection methods can also be used to detect interior defects after the build has been completed, eliminating the need for post-processing microscopy. This is enabled by the real-time retrieval of robot position data through ROS, which can be registered with the predicted quality labels to facilitate location-specific defect identification. Once the defective regions are identified, robotic machining can be applied to remove them. Therefore, the in-situ defect detection strategy sets the foundation for developing a self-adaptive hybrid processing strategy that is capable of enhancing part quality and streamlining the printing process.

## Acknowledgments

This research is funded by the Agency for Science, Technology and Research (A*STAR) of Singapore through the Career Development Fund (Grant No. C210812030). It is also supported by Singapore Centre for 3D Printing (SC3DP), the National Research Foundation, Prime Minister's Office, Singapore under its Medium-Sized Centre funding scheme.

# Appendix A. Acoustic feature correlation matrix

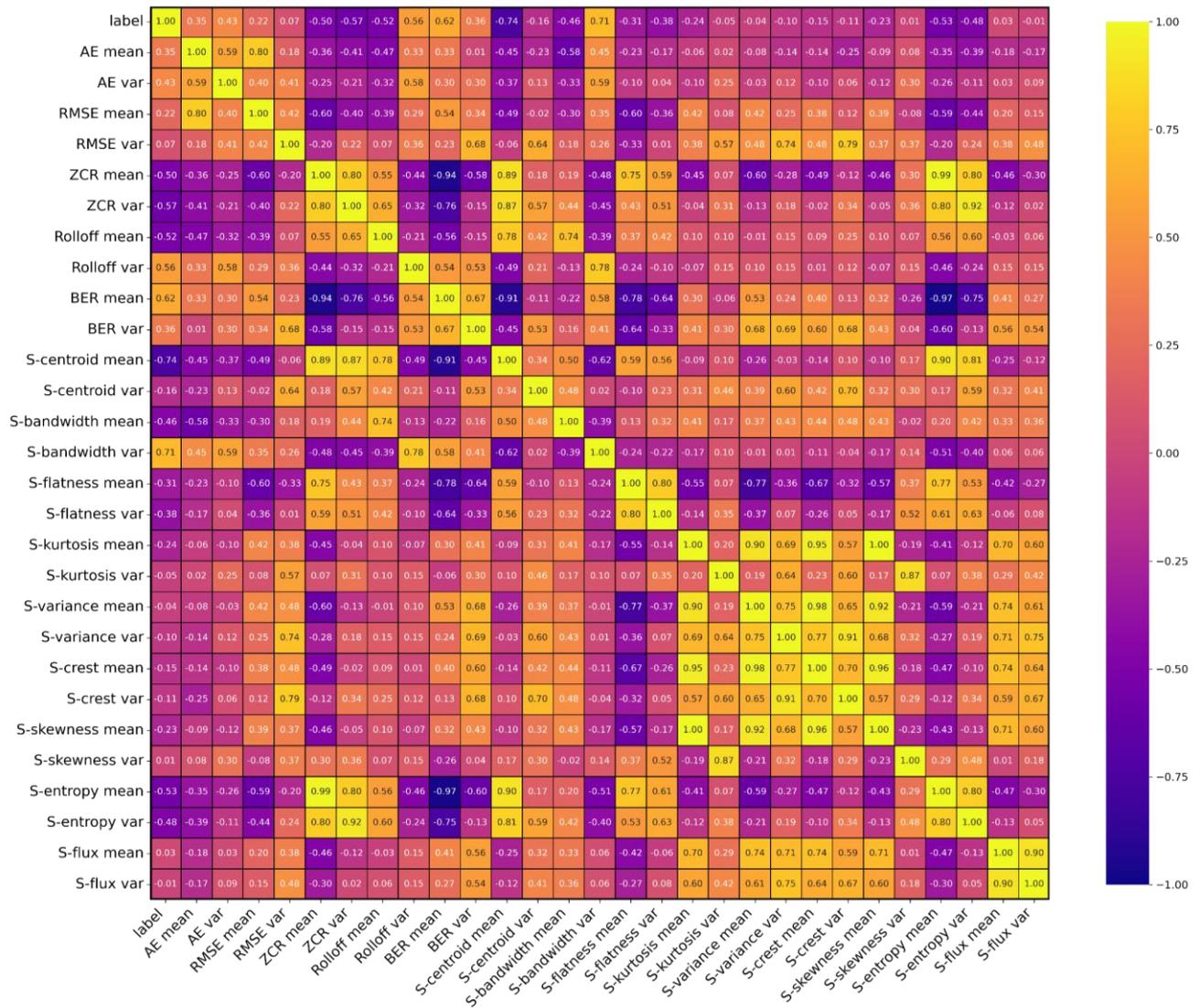

**Figure A1**. Heat map that show correlations amongst numerical acoustic features (time-domain and spectral descriptors) and the output class (i.e., 0 – defect free, 1 - cracks, 2 - keyhole pores).